\input{aipcheck}

\documentclass[
    ,final            % use final for the camera ready runs
  ]
  {aipproc}

\layoutstyle{6x9}

\usepackage{epsfig}
\usepackage{graphicx}
\usepackage{mathptmx}      % use Times fonts if available on your TeX system
\usepackage{latexsym}
\usepackage{bm}
\usepackage{amsfonts}
\usepackage{amsmath}

\def\slashchar#1{\setbox0=\hbox{$#1$}
   \dimen0=\wd0 \setbox1=\hbox{/} \dimen1=\wd1
   \ifdim\dimen0>\dimen1 \rlap{\hbox to \dimen0{\hfil/\hfil}} #1
   \else  \rlap{\hbox to \dimen1{\hfil$#1$\hfil}} / \fi}

\def\tr{{\rm tr}}

\newcommand{\MeV}{\,{\rm MeV}}

\newcommand{\SU}{{\rm SU}}

\newcommand{\Eq}[1]{Eq.~(\ref{eq:#1})}

\newcommand{\vx}{{\bm{x}}}
\newcommand{\vp}{{\bm{p}}}

\newcommand{\ignore}[1]{}

\begin{document}

\title{From  Chiral quark dynamics with Polyakov loop to the hadron resonance gas model~\footnote{Talk given by E. R. A. at XII HADRON PHYSICS, April 22-27 2012, Bento Goncalves, Wineyards Valley Region, Rio Grande do Sul, Brazil}~~~\footnote{Supported by
DGI (FIS2011-24149 and FPA2011-25948) and Junta de Andaluc{\'\i}a
grant FQM-225.}}
%\title{}


\author{\underline{E. Ruiz Arriola}}{
  address={Departamento de F{\'\i}sica At\'omica, Molecular y Nuclear  and
  Instituto Carlos I de F{\'\i}sica Te\'orica y Computacional.  Universidad
  de Granada, E-18071 Granada, Spain}
}
\author{E. Megias}{
address={Grup de F\'{\i}sica Te\`orica and IFAE, Departament de F\'{\i}sica, Universitat Aut\`onoma de Barcelona, Bellaterra E-08193 Barcelona, Spain}
}
\author{L. L. Salcedo}{
  address={Departamento de F{\'\i}sica At\'omica, Molecular y Nuclear and
  Instituto Carlos I de F{\'\i}sica Te\'orica y Computacional. Universidad
  de Granada, E-18071 Granada, Spain}
}

\begin{abstract}
Chiral quark models with Polyakov loop at finite temperature have been
often used to describe the phase transition. We show how the
transition to a hadron resonance gas is realized based on the quantum
and local nature of the Polyakov loop.
\end{abstract}

\maketitle

%%%%%%%%%%%%%%%%%%%%%%%%%%%%%%%%%%%%%%%%%%%%
%% MAINMATTER
%%%%%%%%%%%%%%%%%%%%%%%%%%%%%%%%%%%%%%%%%%%%

\section{Introduction}
\label{sec:introduction}

The cross-over between the chiral symmetry restoration and
deconfinement at a common critical temperature $T_c \sim 200 {\rm
MeV}$ was a first~\cite{Kogut:1982rt,Polonyi:1984zt,Pisarski:1983ms}
and by now firmly established prediction of lattice
QCD~\cite{Aoki:2006we,Bazavov:2011nk}. The transition point is
characterized by two order parameters.  On the one hand the quark
condensate $\langle
\bar q q \rangle$ vanishes smoothly.  On the other hand the Polyakov
$L_T=\langle \tr_c e^{iA_0/T}\rangle /N_c$ loop, where $A_0 $ is a gluon
field in the adjoint representation of the $\SU(N_c)$ gauge group,
jumps smoothly from zero to one signaling the breaking of the center
symmetry $\mathbb{Z}(N_c)$~\cite{Svetitsky:1985ye}. Of course,
$\langle \bar q q \rangle$ and $L_T $ are true order parameters in the
opposite limits of vanishing quark masses (chiral limit) and for
infinitely quark masses (gluodynamics) respectively, while the true
cross-over is defined by inflexion points of both $\langle \bar q q
\rangle$ and $L_T $. The expectation that a phase transition between a
hadronic phase to the quark-gluon plasma phase could be observed on
the laboratory has inspired a wealth of work in recent
years~\cite{Fukushima:2011jc}.

There have been many attempts to model the chiral-deconfinement
cross-over, the main difficulty lies in properly combining the
relevant degrees of freedom for the corresponding order parameter;
while well below $T_c$ hadrons provide a complete basis of states,
much above $T_c$ just quarks and gluons seem the adequate basis.  The
cross-over region seems difficult as it marks the coexistence of both
degrees of freedom. In this contribution we will focus on the low
temperature region of chiral quark models where the crossover is known
to occur at higher temperature.

\section{The hadron resonance gas model}

On general grounds, and because quarks and gluons are confined, one
expects that at sufficiently low temperatures {\it any} observable in
QCD should be represented by the relevant hadronic colour neutral
states.

For the vacuum energy at finite temperature, and more specifically the
trace of the energy momentum tensor, $\Delta(T)=(\epsilon-3p)/T^4$,
the results found on the lattice are expected to be represented by an
interacting gas of low-lying stable hadrons (for light $u$ and $d$
quarks it would just be a gas of $\pi$, $N$ and $\bar N$ states). Most
of the interactions in the scattering region generate resonances (such
as $\rho,\omega,\Delta, etc.)$ which could be represented as narrow
states provided the ratio $\Gamma/M$ is small.  This is consistent
with the large $N_c$ limit expectation that $\Gamma/M =
{\cal O}(N_c^{-1})$ while experimentally 
$\Gamma/M=0.12(8)$~\cite{Arriola:2011en,Masjuan:2012gc} for both
mesons and baryons listed in the Particle Data Group (PDG)
booklet~\cite{Nakamura:2010zzi}. Thus, in the Hadron Resonance Gas
(HRG) model the interactions are represented by a bunch of
narrow-looking resonances whose partition function is given by
\cite{Hagedorn:1984hz, Megias:2009mp,
  Huovinen:2009yb, Ratti:2010kj, Bazavov:2011nk},
\begin{equation}
\log Z = 
-\int \frac{d^3 x \, d^3 p}{(2\pi)^3} \sum_\alpha \zeta_\alpha  z_\alpha g_\alpha 
\log \left( 1 - \zeta_\alpha e^{-\sqrt{p^2+M_\alpha^2}/T} \right) \,,
\label{eq:rgm}
\end{equation}
with $g_\alpha=(2J_\alpha+1)(2T_\alpha+1)$ the degeneracy factor,
$\zeta_\alpha=\pm 1$ for bosons and fermions respectively, $M_\alpha$
the hadron mass and $z_\alpha$ the fugacity. The PDG
states~\cite{Nakamura:2010zzi} saturate lattice
calculations~\cite{Bazavov:2009zn,Borsanyi:2010bp}~\footnote{A
temperature shift of about $T_0=10-20 {\rm MeV}$ is required;
$\Delta_{\rm HRG} (T-T_0)=\Delta_{\rm QCD} (T)$ in
Ref.~\cite{Bazavov:2009zn}.} as also found within a strong coupling
expansion for heavy quarks in \cite{Langelage:2010yn}.

This HRG representation is less obvious for QCD operators involving
just gluon fields. However, we have recently
shown~\cite{Megias:2012kb} that a hadronic representation of the
Polyakov loop is given by
\begin{eqnarray}
L_T = \frac1{N_c} \langle \tr e^{iA_0/T}\rangle \approx \frac{1}{2
N_c}\sum_\alpha g_\alpha e^{-\Delta_\alpha/T},
\label{eq:L_HRG}
\end{eqnarray}
where $g_\alpha$ are the degeneracies and $\Delta_\alpha$ are the
masses of hadrons with exactly one heavy quark (the mass of the heavy
quark itself being subtracted). The comparison with the spectrum with
$u,d,s$ light quarks and one extra heavy quark turns out to be rather
satisfactory and fairly independent on taking charm, bottom or truely
infinite heavy quarks. It is also intriguing since these calculations
might provide a handle on deciding the existence of exotic multiquark
states~\cite{Megias:2012kb}.

These HRG approximations are expected to hold at sufficiently low
temperatures and agreement with lattice data is observed within the
finite lattice uncertainties. In any case, it turns out that many
states are needed to saturate both the trace anomaly and the Polyakov
loop at temperatures below $T_c$ as there are no significant gaps in
the spectrum.

\section{Polyakov-Chiral-Quark models and heavy quarks}
\label{sec:PolyakovCQ}

An effective and phenomenologically successful approach to the physics
of the phase transition is provided by chiral quark models coupled to
gluon fields in the form of a Polyakov
loop~\cite{Meisinger:1995ih,Fukushima:2003fw,Megias:2004hj,Megias:2006bn,Ratti:2005jh,Sasaki:2006ww,Ciminale:2007sr,Contrera:2007wu,Schaefer:2007pw,Costa:2008dp,Mao:2009aq,Sakai:2010rp,Radzhabov:2010dd,Zhang:2010kn}.
Because most often the venerable Nambu--Jona-Lasinio model has been
used, these models are referred to as PNJL models.  In this
contribution we discuss how PNJL models may be represented as a HRG at
low T.  Many works remain within a mean field approximation and assume
a {\it global} Polyakov loop. As we have repeatedly criticized in our
previous
works~\cite{Megias:2004hj,Megias:2006bn,Megias:2005qf,Megias:2006df}
this raises the theoretical problem of the undesirable ambiguity of
group coordinates on the one hand, but also the practical problem that
the adjoint-representation provides a non-vanishing value for the
Polyakov loop, contradicting lattice simulations. These difficulties
may be overcome~\cite{Megias:2004hj,Megias:2006bn} by recognizing the
local and quantum nature of the Polyakov loop.

We start out from the partition function
motivated in~\cite{Megias:2004hj,Megias:2006bn,Megias:2005qf,Megias:2006df}
\begin{equation}
Z= \int D \Omega \, e^{-S(T, \Omega)} \,,
\label{eq:3}
\end{equation}
where $\Omega = e^{i A_0/T}$ and $D \Omega$ is the invariant
$\SU(N_c)$ Haar group integration measure, for each $\SU(N_c)$ variable
$\Omega(\bm x)$ at each point $\vx$.  Here the action is
\begin{equation}
S (T,\Omega) =  S_q (T,\Omega) +  S_G (T,\Omega) 
.
\end{equation}
The fermionic contribution depends on the quarks (and anti-quarks) is
obtained from the corresponding fermion determinant. 
Assuming mass-degenerated quarks for simplicity reads
\begin{eqnarray}
S_q (T,\Omega) =  -2 N_f  \int \frac{d^3
  x d^3 p}{(2\pi)^3} \bigg( 
 \tr_c \log \big[ 1+\Omega ({\bm x}) \, e^{-E_p/T}\big] +  \tr_c \log \big[ 1+\Omega^\dagger ({\bm x}) \,e^{-E_p/T}\big] 
\bigg) 
.
\label{eq:quark-action}
\end{eqnarray}
Here $E_p = \sqrt{\vp^2+M^2}$ is the energy and $M$ the
constituent quark mass. As one can see the diagonal part of the
Polyakov loop corresponds to consider chemical potentials for
different color species. Large color gauge invariance is implemented
by just averaging over group elements.

The partition function character of the Polyakov loop can be
appreciated considering a system with $N_f$ dynamical quarks and an
extra putative heavy quark (not antiquark) of mass $m_H$ at rest
located at a fixed point and with fixed spin and colour $a=1, \dots
N_c$.  From Eq.~(\ref{eq:quark-action}) the change in the effective
action is
\begin{eqnarray}
S_{q} (N_f+ 1) - S_q (N_f) &=& -2 \log (1+ \Omega_{aa} e^{-E_h/T} )  
\approx -2 e^{-m_H /T} \Omega_{aa} 
\end{eqnarray}
yielding  the partition
function
\begin{eqnarray}
\frac{Z(N_f+ 1)}{Z(N_f)} &=&  1+\langle \Omega_{aa} \rangle 2 e^{-m_H/T}  
= 1+\frac1{N_c}\langle \tr_c \Omega \rangle 2 e^{-m_H/T}  )
\end{eqnarray}
after averaging over color degrees of freedom implied by
$D\Omega$. Thus we get 
\begin{eqnarray}
\frac1{N_c}\langle \tr_c \Omega \rangle = \lim_{m_H \to \infty} \frac12 \left[\frac{Z(N_f+ 1)}{Z(N_f)} -1 \right] e^{m_H/T}
\end{eqnarray}
If we use the HRG, Eq.~(\ref{eq:rgm}), to evaluate the r.h.s.  we
reproduce the HRG result, Eq.~(\ref{eq:L_HRG}) for the Polyakov loop,
providing confidence on the assumed coupling to quarks.

The action $S_G(\Omega)$ would follow from gluodynamics but it is
exponentially suppressed at low temperatures and the distribution of
$\Omega$ locally coincides with the Haar measure. A convenient model
to account for Polyakov loop correlations at different points and
compatible with group integration at equal points~\cite{Creutz:1978ub}
is~\cite{Svetitsky:1985ye}
\begin{equation}
\langle \tr_c \Omega (\vx) \; \tr_c
\Omega^{-1} (\bm{y}) \rangle_{S_G} = e^{-\sigma |\vx-\bm{y}| /T} \,, 
\label{eq:corr-func}
\end{equation} 
with $\sigma$ the string tension, including the correct screening of
the color charge at large distances. This defines a correlation length
and independent confinement domains with volume
$V_\sigma \equiv \frac{8 \pi T^3}{\sigma^3}$ which describe the
cross-over between deconfinement and chiral symmetry
restoration~\cite{Megias:2004hj,Megias:2006bn,Megias:2006df}.

Using these assumptions let us now consider the calculation of the partition
function at low temperatures where to lowest
order one gets contributions of just $q \bar q$ states
\begin{eqnarray}
\log Z = (2 N_f)^2  
\int \frac{d^3 x_1 d^3 p_1}{(2\pi)^3} 
\int \frac{d^3 x_2 d^3 p_2}{(2\pi)^3}  e^{-E_1/T} e^{-E_2/T}  \langle \tr_c \Omega (\vx_1) \tr_c \Omega^\dagger (\vx_2)\rangle_{S_G} 
 + \cdots 
\end{eqnarray}
which using 
Eq.~(\ref{eq:corr-func}) can be rewritten as 
\begin{equation}
\log Z =  (2N_f)^2  
\int 
\frac{d^3 x_1 d^3 p_1}{(2\pi)^3} 
\frac{d^3 x_2 d^3 p_2}{(2\pi)^3} e^{-H(\vx_1,\vp_1;\vx_2,\vp_2)/T} + \cdots \,.
\label{eq:17}
\end{equation}
We recognize the {\it classical} partition function of a $\bar q q$
system with a Hamiltonian
\begin{equation}
H(\vx_1,\vp_1;\vx_2,\vp_2)= \sqrt{\vp_1^2+M^2}+\sqrt{\vp_2^2+M^2}+
\sigma r_{12} \,.
\label{eq:salp-2}
\end{equation}
Separating CM and relative motion, direct integration gives at low $T$ values
\begin{eqnarray}
\log Z \approx (2 N_f)^2 V \int d^3 x \, e^{-\sigma r /T}
\left[\int \frac{d^3 p}{(2\pi)^3} e^{-E_p/T} \right]^2 
\approx (2 N_f)^2 V V_\sigma 
\left(\frac{M T}{2 \pi} \right)^3
e^{-2 M/T}
.
\label{eq:13}
\end{eqnarray}
Similarly, for the Polyakov loop (heavy quark located at $\vx_0$) we get 
\begin{eqnarray}
L_T &=& 2 N_f
\int \frac{d^3 x \, d^3 p}{(2\pi)^3} 
e^{-E_p/T}
\frac{1}{N_c}\langle\tr_c\Omega(\vx_0) \,\tr_c\Omega^{-1}(\vx)\rangle_{S_G}
+\cdots
\label{eq:Llow}
\end{eqnarray}
whence one can also rewrite the expression as 
\begin{eqnarray}
L(T) = \frac{2 N_f}{N_c} \int \frac{d^3 x \, d^3 p }{(2\pi)^3} e^{-H(\vx,\vp)/T}
+\cdots .
\end{eqnarray}
This corresponds to the {\it classical} partition function of the one-quark
Hamiltonian 
\begin{equation}
H(\vx,\vp)= \sqrt{\vp^2+M^2} +  \sigma r \,.
\label{eq:salp-1}
\end{equation}
where after integration 
\begin{eqnarray}
L(T) = \frac{N_f}{N_c} V_\sigma 
\frac{M^2 T}{\pi^2} K_2 \left( \frac{M}{T}\right) 
+\cdots \approx
\frac{2 N_f}{N_c} V_\sigma
\left(\frac{M T}{2 \pi} \right)^{3/2} e^{-M/T}
.
\label{eq:15}
\end{eqnarray}
As it was shown in our previous works these rules provide a
satisfactory phenomenological description of the chiral-deconfinement
cross-over observed in lattice calculations. The previous lowest order
approximations do indeed reproduce the more sophisticated results up
to $T \sim 0.75 T_c$~\cite{Megias:2005qf}.

\section{Quark-Hadron Duality at finite temperature}
\label{sec:Quantizing}

While the previous formulation reproduces lattice results in a
satisfactory manner at $T_c \approx 250 {\rm MeV}$, within a hadronic
phase it should be possible to express {\it all observables} in terms
of purely hadronic properties, a feature absent in the model.  To
improve on this from Eqs.~(\ref{eq:17}) and (\ref{eq:salp-2}) we apply
standard quantization rules to the relativistic two-body
quantum-mechanical problem, which in the CM system ($\vp_1+\vp_2=0$)
corresponds to solve a Salpeter equation for scalar particles
\begin{equation}
\left( 2 \sqrt{\vp^2 + M^2} + \sigma r \right) \psi_n = M_n \psi_n \,.
\label{eq:salp-2-CM}
\end{equation}
In a more elaborated treatment a two-body Dirac equation should be
obtained.  Note that at any stage our approach is Pauli-principle
preserving at the quark level. \Eq{salp-2}, or its Dirac version,
describes the interaction of $q\bar{q}$ pairs to form mesons, each
meson with $(2N_f)^2$-fold degeneracy.  The crucial point here is to
keep track of the number and labeling of states contributing to the
sum in \Eq{17}. Thus after quantization and implementation of
relativistic invariance, we get
\begin{equation}
\log Z =  \int \frac{d^3xd^3 p}{(2\pi)^3} \sum_{\alpha} g_\alpha  
e^{-\sqrt{p^2+M_\alpha^2}/T}
+\cdots
\end{equation}
Here $\vx,\vp$ represent the former CM coordinates of the $q\bar{q}$
pair. Note that the result holds even if quark masses are not degenerated.

The expression obtained nicely reproduces the first bosonic term in
the expansion of the RGM in \Eq{rgm}. We are assuming that the pion is
also contained in the sum although the dynamics leading to such a
state is not literally determined by the specific Hamiltonian in
Eq.~(\ref{eq:salp-2}).  The extension of the previous result to
multiquark mesonic and baryonic states will be presented
elsewhere~\cite{Megias2012}. 

The argument can be extended to the Polyakov loop by re-quantization
of the heavy-light Hamiltonian Eq.~(\ref{eq:salp-1}) or its Dirac
counterpart yielding the eigenvalues $\Delta_\alpha$ and reproducing
indeed the HRG form, Eq.~(\ref {eq:L_HRG}), discussed in
Ref.~\cite{Megias:2012kb} as required by quark-hadron duality. This
constitutes our main insight which suggests re-analyzing these models
after incorporating the connection to the HRG~\cite{Megias2012}.

The original Polyakov loop models suggested that $L_T$ and $Z$ (and
hence $\langle \bar q q \rangle_T$) are closely intertwined through
their exponential suppression at low temperatures, controlled by the
constituent quark mass (Eqs.~(\ref{eq:13}) and
(\ref{eq:15})). However, in the hadronized version, the suppression
depends on two not directly related hadron masses, namely, the pion
for $Z$ and the lightest meson with a heavy quark for the Polyakov
loop. It is also noteworthy that the prefactors, namely, the powers of
$T$ present in the original model for both $Z$ and $L(T)$, are removed
by the quantization. The origin of these prefactors was the relative
motion in the $q\bar{q}$ pair, which produces a continuous spectrum in
the classical case, but yields a discrete spectrum after quantization,
typical of quantum bound states. Finally, the independence between
deconfinement and chiral symmetry restoration corresponding to crossed
correlator $\langle \bar q q \, \tr
e^{iA_0/T} \, \rangle_T \sim \partial L / \partial m_q \approx 0 $
becomes evident at low $T$ due to the weak dependence of the Polyakov
loop and thus of the static spectrum $\Delta_\alpha$ on the current
quark mass. These features are also seen in lattice
calculations~\cite{Bazavov:2011nk}.

\bibliographystyle{aipproc}   % if natbib is available
%\bibliography{refs}

\end{document}